**ANALYTICAL REPORT**

# Science vs Propaganda: The case of Russia

**Alex Plastun**

**Sumy State University, Ukraine**

**Inna Makarenko**

**Sumy State University, Ukraine**

**University of Helsinki, Finland**

**Tetiana Hryn'ova**

**CNRS, France**

*This note highlights how Russia uses the international academic sphere—including scientometric databases, international publishers, and international organizations—as a propaganda tool to legitimize its appropriation of Ukrainian territories.*

## Science vs Propaganda: The case of Russia

Following unprovoked and unjustified aggression of the Russian Federation against Ukraine, numerous sanctions targeting the Russian Federation, Russian companies, and Russian individuals across various sectors, including the military, economy, culture, sports, and science have been implemented. Currently, over 22,000 different sanctions are active against Russia, the highest number in human history. Russia is now more than three times more sanctioned than Iran and more than five times more than Syria or North Korea (https://www.castellum.ai/russia-sanctions-dashboard).

The application of sanctions in the field of science is particularly controversial due to a myth that science is outside of politics. Despite the ongoing debate and opposition (both silent and vocal) against sanctions, many measures have been adopted. These include funding restrictions, halting collaborations with Russian institutions and closing joint projects, imposing restrictions and limitations on equipment and reagents, and scientists have refused to participate in scientific conferences and international events organized by Russian institutes.

Despite the above, the existing sanctions against Russian science are insufficient, allowing the Russian scientific community to spread its propaganda through international academic journals and papers. The most commonly observed cases include:

- ***Appropriation of Ukrainian territories***: This involves providing affiliation with the Russian Federation to various Ukrainian regions in editorial boards or academic papers and books.
- ***Spread of Russian propaganda narratives to support its aggression against Ukraine***: This appears in academic papers that justify and support Russia's actions.
- ***Support and justification of aggression by Russian academic organizations***: Notable examples include the so-called Russian Rectors' letter and statements by the Russian Academy of Sciences and its Presidium. The Academic Council of Lomonosov Moscow State University also issued a statement in support of the war, along with individual academics from institutions like Saint Petersburg State University and others.

This report focuses on the first case - the legitimization of the annexation of occupied Ukrainian territories by the Russian Federation - within the international academic sphere.

## Appropriation of Ukrainian Territories

One of the most common methods of propaganda by Russia in the academic journals is providing affiliation with the Russian Federation to Ukrainian regions, either through institutions of authors or editorial board members, or conference locations or in abstracts and references. This practice is wide-spread on the websites of most international publishers as detailed below.

### Study 1: Authors, Journals in the Scopus database and Springer journals

Our investigation includes data from the Scopus database and Springer journals. The most typical illustrations of this method can be found in the screenshots provided in Appendix A (Scopus) and Appendix B (Springer Nature).

The Scopus platform not only allows us to identify individual cases but also provides the means to extract statistics related to the frequency of such situations, proving that this is a systematic issue and not an isolated incident.

## Search Algorithm

The search algorithm was as follows: a Ukrainian city of interest is queried. For example, “Donetsk”. This allows us to see the total number of publications affiliated with Donetsk. Next, we filter it with the limitation of the territory “Russian Federation”. This allows us to find the number of publications where Donetsk is affiliated with Russia. In the resulting list an additional search is performed for a mention of “Ukraine” in the affiliation to make sure that the resulting combination is not spurious due to the presence of another institute based in Russia in the author list. Such cases are manually removed from the count.

Here is an example of a search query link for the case of Donetsk: Scopus Search Query for Donetsk:

https://www.scopus.com/results/results.uri?sort=plf-f&src=s&st1=Donetsk&sid=9f3aa224b52d0904fc105b3f9ec2f63e&sot=b&sdt=b&sl=18&s=AFFILCITY%28Donetsk%29&origin=searchbasic&editSaveSearch=&yearFrom=Before+1960&yearTo=Present&sessionSearchId=9f3aa224b52d0904fc105b3f9ec2f63e&limit=10

The following Ukrainian territories were used as objects of analysis:

- Donetsk
- Lugansk
- Sevastopol
- Simferopol
- Yalta
- Kerch
- Crimea
- Mariupol

This is not an exhaustive list of Ukrainian territories marked as Russian (other examples include Feodosia, Yevpatoria, Makeevka, Alchevsk, and many others), but for illustrative purposes, this selection is more than sufficient.

## Results

The results of our analysis are presented in Table 1. These findings highlight a systematic issue of misrepresenting Ukrainian territories as part of the Russian Federation in academic publications. The fraction of publications affected depends on the duration of the occupation by the Russian Federation and the claimed annexation date[1]. It is over 90% for the cities in Crimea, which were temporarily occupied and claimed to be annexed by Russia in 2014, versus 50% for Donetsk and Lugansk temporarily occupied in 2014 but claimed to be annexed only in 2022. For Mariupol (temporarily occupied by Russia since 2022), it remained zero in 2022 rising to 4% in 2023, which is an unfortunate trend. An interesting case is Kherson, which was briefly occupied for a few months in 2022. No publications affiliated Kherson with the Russian Federation in 2022 and 2023, but in 2024 (two years after it was liberated by Ukraine) two papers published by edpScience appeared in Scopus [1,2]. This is how silent support works. In ten years, we might have a picture similar to Crimean cities in all temporarily occupied Ukrainian territories.

**Table 1: Ukrainian Territories Marked as Russian in Scopus**

| Ukrainian territory/city | 2022 | 2023 | % in 2022 | % in 2023 |
|---|---|---|---|---|
| Donetsk | 154 | 177 | | |
| Donetsk is affiliated with the Russian Federation | 16 | 115 | 10% | 65% |
| Lugansk | 38 | 46 | | |
| Lugansk affiliated as Russian Federation | 1 | 23 | 3% | 50% |
| Sevastopol (Crimea) | 575 | 633 | | |
| Sevastopol affiliated as Russian Federation | 556 | 625 | 97% | 99% |
| Simferopol (Crimea) | 379 | 444 | | |
| Simferopol is affiliated with the Russian Federation | 348 | 413 | 92% | 93% |
| Yalta (Crimea) | 69 | 81 | | |
| Yalta is affiliated with the Russian Federation | 60 | 75 | 87% | 93% |
| Kerch (Crimea) | 27 | 29 | | |

[1] According to the so-called Constitution of the Russian Federation, both Crimea, Lugansk, Donetsk, Mariupol and Kherson are claimed to be Russian. This annexation is not recognized by the international community. ["General Assembly Adopts Resolution Urging Russian Federation to Withdraw Its Armed Forces from Crimea, Expressing Grave Concern about Rising Military Presence". United Nations. 17 December 2018. Archived from the original.]

| | | | | |
|---|---|---|---|---|
| Kerch is affiliated with the Russian Federation | 19 | 29 | 70% | 100% |
| Crimea | 11 | 15 | | |
| Crimea is affiliated with the Russian Federation | 10 | 11 | 91% | 73% |
| Mariupol | 100 | 46 | | |
| Mariupol is affiliated with the Russian Federation | 0 | 2 | 0% | 4% |

Source: elaborated by authors (in-built Scopus instruments).

Upon our inquiry to Scopus regarding this situation, their response was as follows:

*"When displaying third party data in the product, Scopus does not make changes to the data and remains true to the original source and the information and data is shown as such."* In other words, Scopus does not take responsibility for the information provided in its database. This stance implies that Scopus will continue to provide its platform for Russian propaganda.

**Scientific entities at the source of mis-affiliations**

Main organizations which propagate those mis-affiliations in the Scopus database are detailed in Table 2. At the heart of the problem is that those institutions, which provide affiliations in the occupied territories of Ukraine were created by Russia using the captured Ukrainian scientific infrastructure and often named similarly or the same as the existing Ukrainian institutions-in-exile. This aims to reassure the international scientific community as to the legitimacy of those entities. Russia works to integrate those impostor organizations into the international scientific community through their scientific research (international partnerships[2], projects and publications), to the detriment of the legal Ukrainian scientific institutions. Many of the institutes on the list have been illegally appropriated by the Russian Academy of Sciences (RAS).

In addition, existing institutions from the Russian Federation were encouraged to move or to open branches in the occupied territories to facilitate their integration into the Russian scientific community. One such example shown in Table 2 is Sochi-based Federal State Budgetary Science Enterprise the Institute of Natural and Technical Systems which moved to occupied Sevastopol in 2014 and appropriated two subdivisions of Marine Hydrophysical Institute of NANU:

[2] For example, in April 2024 the University of Lucknow signed a Memorandum of Understanding with the Institute of Biology of the Southern Seas, the oldest "Russian" institute of science which was stolen from Ukraine.

Special Design and Technology Bureau and Science and Production Center Ecosi-Hydrophysics. It currently publishes under the name of Institute of Natural and Technical Systems.

**Table 2 Main organizations which are used to affiliate Ukrainian territories to Russia, together with a number of affected publications listed by Scopus for 2023.**[3]

| **Impostor Institutes created by Russia (different spellings exist)** | **Original Ukrainian Institutes** | **Changed location** | **Publications in 2023** | **Locations** |
|---|---|---|---|---|
| V.I. Vernadsky Crimean Federal University (2014), under sanctions | V.I. Vernadsky Taurida National University currently in Kyiv | Y | 376 (355/21) | Simferopol, Yalta, Crimea |
| Sevastopol State University (2014), under sanctions | Sevastopol National Technical University; Sevastopol City Humanitarian University; Sevastopol National University of Nuclear Energy and Industry; Sevastopol Faculty of Maritime Transport and Sevastopol Maritime College of the Kyiv State Academy of Water Transport (located in Kyiv); Sevastopol branches of South Ukrainian National Pedagogical University (located in Odesa) and Donetsk National University of Economics and Trade (located in Kryvyi Rih) | N | 223 | Sevastopol, Crimea |
| Marine Hydrophysical Institute of RAS (2014) | Marine Hydrophysical Institute of National Academy of Sciences of Ukraine (NANU) | N | 222 | Sevastopol, Crimea |
| A. O. Kovalevsky Institute of Biology of the Southern Seas of RAS (2014) | A. O. Kovalevsky Institute of Biology of the Southern Seas of NANU renamed in 2014 into Institute of Marine Biology of NANU, currently in Odesa | Y | 185 | Sevastopol, Crimea |

[3] Notes: The approximate date of impostor institute creation is given in brackets, if known, together with its current sanctions status. Last two columns show the number of publications in Scopus in 2023 together with their locations which they affiliate to the Russian Federation.

| | | | | |
|---|---|---|---|---|
| Donetsk National Medical University/Donetsk State Medical University (2014) | Donetsk National Medical University of NANU currently in Kropyvnytskyi | Y | 32 (19/13) | Donetsk |
| Nikitsky Botanical Gardens—National Scientific Center of RAS (2014) | Nikitsky Botanical Gardens - National Scientific Center of National Academy of Agrarian Sciences of Ukraine | N | 29 | Yalta, Crimea |
| Donetsk State University/Donetsk National University (2014) | Vasyl' Stus Donetsk National University which was moved to Vinnytsia | Y | 28 (17/11) | Donetsk |
| Kerch State Maritime Technological University (2014), under sanctions | Kerch State Maritime Technological University | N | 25 | Kerch, Crimea |
| All-Russian National Research Institute of Winegrowing and Winemaking "Magarach," RAS (2014), under sanctions | Institute of grapes and wine "Magarach" of the National Academy of Agrarian Sciences of Ukraine | N | 25 | Yalta, Crimea |
| A. A. Galkin Donetsk Institute for Physics and Engineering (2016) | A. A. Galkin Donetsk Institute for Physics and Engineering of NANU, currently in Kyiv | Y | 24 | Donetsk |
| Institute of Archaeology of Crimea (2014) | Crimean branch of the Institute of Archaeology of NANU, currently in Odesa | Y | 24 | Simferopol, Crimea |
| Research Institute of Agriculture of Crimea (2014), under sanctions | Crimean Agrotechnological University, currently in Kyiv | Y | 22 | Simferopol, Crimea |
| Institute of Natural and Technical Systems (originally Sochi-based Federal State Budgetary Science Enterprise the Institute of Natural and Technical Systems, 2014), under sanctions | Special Design and Technology Bureau and Science and Production Center Ecosi-Hydrophysics of Marine Hydrophysical Institute of NANU | n/a | 21 | Sevastopol, Crimea |
| Donetsk National Technical University (2014) | Donetsk National Technical University currently in Lutsk | Y | 15 | Donetsk |
| Lugansk State Pedagogical University (2020) | Taras Shevchenko National University of Luhansk currently in Poltava | Y | 9 | Lugansk |
| Crimean Astrophysical Observatory of RAS (2015) | Research Institute "Crimean Astrophysical Observatory", Ministry of Science of Ukraine | N | 9 | Nauchny, Katsiveli, Simeiz, Yalta, Crimea |

**Publishers' Role**

Databases, like Scopus, do not correct materials provided to them by journals (although they can decide if to host the illegitimate information on their websites or not). The issue is that journals choose to accept publications with problematic affiliations. Table 3 breaks down the publications in Scopus for the year 2023 shown in Table 1 by publisher. Notably over 30% of the affiliations to Russia per city are propagated through the international publishers, heavily dominated by Springer Nature/Pleiades (Germany/USA), followed by MDPI (Switzerland), EDP Science (France), IEEE (USA) and Elsevier (Netherlands). Individual publications appeared in ASV Publishing, CSIRO, MM Publishing, Science Press, Wolters Kluwer Medknow Publications and World Scientific.

It is not surprising that Springer Nature/Pleiades Publishing partnership leads the list: together with Pleiades subsidiary Allerton Press, they are the leading global provider of English language journals from the former USSR, publishing over 200 such English-language journals (emanating from over 270 local language journals) on the Springer website.

**Table 3 Number of publications by Publisher in Scopus in 2023, affiliating Ukrainian cities to the Russian Federation.**

| Scopus, 2023 | Sevastopol | Simferopol | Donetsk | Yalta | Kerch | Lugansk | Crimea | Mariupol |
|---|---|---|---|---|---|---|---|---|
| Total | 625 | 413 | 115 | 75 | 29 | 23 | 11 | 2 |
| Springer Nature, Germany/ Pleiades, USA | 194 | 64 | 40 | 15 | 7 | 10 | 9 | |
| MDPI, Switzerland | 65 | 25 | 5 | 12 | 2 | 1 | | 2 |
| EDP Science, | 43 | 32 | 4 | 11 | 8 | 1 | | |

| | | | | | | |
|---|---|---|---|---|---|---|
| France | | | | | | |
| IEEE, USA | 37 | 8 | 3 | 6 | 7 | 1 |
| Elsevier, Netherlands/Saunders, USA | 24 | 9 | 3 | 0 | 1 | 0 |
| SPIE, France | 13 | | | | | |
| AIP Publishing, USA | 11 | 14 | | 4 | 1 | 1 |
| Wiley, USA | 10 | 1 | | | | |
| Taylor and Francis, UK | 3 | | | | | |
| EcoVector, Russia | 3 | 2 | | | | |
| Frontiers, Switzerland | 3 | | | | | |
| IOP, UK | 1 | 1 | | 3 | | |
| Cambridge, UK | | 4 | | | | |
| Bentham, UAE | | 2 | | | | 1 |
| Science and Innovation Publishing House, USA | | 5 | 1 | 3 | | |

In Appendix C, you can find a list of journals published by Springer Nature where we have detected cases of Ukrainian territories being marked as Russian. This is not the whole list but a selection of examples. We stopped at the letter “P” and explored only journals with international members on the editorial board. Even with these limitations, we identified 50 journals. As can be seen, this is not a coincidence but a system. The number of cases is much higher because there are journals with dozens of instances of Ukrainian territories marked as Russian.

Another major source of problematic contributions is proceedings of conferences held in Russia. These were published by all major publishers, e.g IOP, PoS/SISSA, AIP, edpScience[4], etc. Some of them have continued to publish not only individual contributions but also proceedings of conferences held in Russia even after the full scale invasion. For example:

- edpScience BIO Web of Conferences published 3 in 2022, 5 in 2023, 3 in 2024;
- PoS/SISSA published two in 2022 only: DLCP2022 and MUTO2022;
- AIP Publishing: ICER 2021 (2022), SMARTICAE (2023), Oil and Gas Engineering (2023), MIP: Engineering-IV-2022 (2024), MIST: Aerospace-V (2024), etc.

This list is not exhaustive.

Below are a few selected examples of contributions to those conferences which affiliate of Ukrainian territories with Russia:

- APEC-IV-2021: 2022 *IOP Conf. Ser.: Earth Environ. Sci.* **990** 012024: Ukrainian city of Sevastopol is affiliated with the Russian Federation
- MIP: Engineering-IV-2022 *AIP Conf. Proc.* 3021, 070022 (2024): where Ukrainian city of Sevastopol is marked as Russia;
- SMARTICAE: *AIP Conf. Proc.* 2910, 020020 (2023) where Ukrainian city of Simferopol is marked as the Russian Federation;
- MIST: Aerospace-V 2023 *AIP Conf. Proc.* 3102, 020035 (2024) where Ukrainian city of Makeevka is marked as Russia.

The issue spans various disciplines, including Chemistry, Physics, Astronomy, Biology, Engineering, Mathematics, Entomology, Physiology, Medicine, and Ichthyology. Traces of Russian propaganda can be found everywhere in the academic sphere.

Some publishers are aware of potential issues, and highlight them in some of their publications, as noted in Table 4 below. Unfortunately these are not applied systematically neither in the case of those publishers nor across the industry.

**Table 4 Examples of publisher notes in articles with illegitimate author affiliations**

| Publisher | Source | Comment |
|---|---|---|
| Springer, Germany | *Russ. Meteorol. Hydrol.* **48**, 871–878 (2023), | "Allerton Press remains neutral with regard to |

[4] Over 30 conferences published in BIO Web of Conferences only between 2017 and 2021.

| | | |
|---|---|---|
| | https://link.springer.com/article/10.3103/S1068373923100060 | jurisdictional claims in published maps and institutional affiliations.” |
| IOP Publishing, UK | *Astrophysical J.* **891** 170 (2020), https://iopscience.iop.org/article/10.3847/1538-4357/ab765d | “While the AAS journals adhere to and respect UN resolutions regarding the designations of territories (available at http://www.un.org/press/en ), it is our policy to use the affiliations provided by our authors on published articles.” |

It is worth mentioning the example of a responsible position taken by the international publisher Brill regarding the Russian journal “Scrinium.” Right after the start of Russia's full-scale invasion of Ukraine, Brill contacted the chief editor of “Scrinium” directly to clarify his position regarding the war and relationship with the RAS. After successfully ensuring compliance, Brill decided to continue publishing “Scrinium.” Additionally, Brill ensures that it does not accept funding from the Russian Federation.

**Study 2: Editorial Boards**

During our investigation, we identified several instances where Ukrainian cities were incorrectly marked as part of the Russian Federation in editorial boards. Below are a few examples:

- ***Journal “Russian Journal of Coordination Chemistry”*** (published by Springer). Editorial member Victor F. Shul’gin is affiliated with the Ukrainian city of Simferopol, but the journal lists it as part of the Russian Federation.
- ***Journal “Water Resources”*** (published by Springer). Editorial member Sergey K. Konovalov is affiliated with the Ukrainian city of Sevastopol, but the journal lists it as part of the Russian Federation.
- ***Journal “Construction Materials and Products”*** (indexed by Scopus): Editorial board member Oleg Nikolaevich Zaytsev is affiliated with the Ukrainian city of Simferopol, but the journal lists it as part of the Russian Federation;
- ***Journal “Ecosystem Transformation”*** (indexed by Scopus): Associate Editor Irina I. Rudneva is affiliated with the Ukrainian city of Sevastopol, but the journal lists it as part of the Russian Federation.

- ***Journal "Arkheologiia Evraziiskikh Stepei"*** (indexed by Scopus): Executive Editor Sergei G. Bocharov is affiliated with the Ukrainian city of Sevastopol, but the journal lists it as part of the Russian Federation. The same misrepresentation applies to Editorial Board member Vladimir P. Kirilko.
- ***Journal "Vestnik Vosstanovitel'noj Mediciny"*** (indexed by Scopus): Editorial board member Vladimir Vladimirovich Ezhov is affiliated with the Ukrainian city of Yalta, but the journal lists it as part of the Russian Federation.

***Study 3: Conferences held in the occupied territories of Ukraine***

In the past ten years, a number of international publishers accepted and spread proceedings of conferences held in the occupied territories of Ukraine (which are advertised on their websites as part of the Russian Federation). Here are a few examples:

- Yearly International Scientific Workshops in Memory of Professor V.P. Sarantsev "Problems of Colliders and Charged Particle Accelerators" are organized by the Joint Institute of Nuclear Research (Russia) in the occupied Alushta, Crimea. The proceedings for 2017 were published by Springer in the JINR-run journal Physics of Particles and Nuclei Letters Volume 15, Issue 7 (2018).
- The International Conference on Modern Trends in Manufacturing Technologies and Equipment (ICMTMTE) was regularly held in the occupied Sevastopol, Crimea. Its proceedings for
  - 2017-2019 were published in MATEC Web of Conferences Volume 129 (2017), Volume 224 (2018) and Volume 298 (2019) by EDPScience (which has as partners French Physics Society, French Chemical Society, French Optics Society, etc);
  - 2020 were published in 2020 *IOP Conf. Ser.: Mater. Sci. Eng.* **971** 011001 by the IOPScience (the publishing company of the Institute of Physics from the UK);
  - 2021 were published in *AIP Conf. Proc.* 2503, 010001 (2022) in October 2022, half a year after the start of the full-scale invasion, by AIP Publishing (subsidiary of the American Institute of Physics, USA).

- The International Scientific-Practical Conference "Modern Trends of Science, Innovative Technologies in Viticulture and Winemaking" has been regularly held in the occupied Yalta, Crimea. Its proceedings for 2021-2023 were published BIO Web of Conferences, Volume 39 (2021), BIO Web of Conferences, Volume 53 (2022) and BIO Web of Conferences, Volume 78 (2023) by EDPScience. Two of them were published after the start of the full-scale invasion.

### *Study 4: Expanding the Issue Beyond Scopus database*

All the key issues discovered in our study for the publications listed in the Scopus database are also true for other participants in the academic publication industry, including:

- Web of Science / Clarivate
- International organizations (e.g., ISSN)
- Repositories and archives of preprints and academic papers (e.g. arXiv, SSRN, inspireHEP, etc)
- Academic Social Nets (e.g. ResearchGate)

#### International organizations: the case of the ISSN

The ISSN, an international organization responsible for the registration of academic journals, has decided not to adhere to ISO 3166. As a result, it provides registration to Russian journals situated in occupied Ukrainian territories (Crimea, Donetsk, and Lugansk regions). No compliance measures have been taken, and no sanctions have been imposed against these journals for spreading propaganda.

#### SSRN

The Social Science Research Network (SSRN) is a repository for preprints devoted to the rapid dissemination of scholarly research in social science. It currently provides no compliance and affiliates Ukrainian territories with Russia, e.g. in papers and for authors (in these examples Ukrainian city of Sevastopol is marked as part of the Russian Federation). It also directly spreads Russian propaganda as discussed in Case 5.

#### Arxiv/inspireHEP

arXiv is a free distribution service and an open-access archive for nearly 2.4 million scholarly articles. Below are a few examples of cases when Ukrainian territories are marked as Russian in either references or author affiliations:

- https://arxiv.org/pdf/2108.11166, where Ukrainian city of Sevastopol is marked as Russian Federation in author affiliation
- https://arxiv.org/pdf/2311.10570, https://arxiv.org/pdf/2311.14287, where Ukrainian city of Alushta is marked as Russian Federation in references

- https://arxiv.org/pdf/2203.09956, https://arxiv.org/pdf/2310.20422, https://arxiv.org/pdf/2207.09544, where Ukrainian city of Simferopol is marked as Russian Federation in author affiliations

Although it is not possible to find such articles easily through arxiv search feature, some additional cases have been identified through inspireHEP database search for "Crimea", "Sevastopol" and "Simferopol", in which over 70 publications with mis-affiliations can be found. Since inspireHEP covers only a limited number of categories in arXiv, the total number of problematic entries present could be significantly higher.

**Academic Social Networks (the case of ResearchGate)**

ResearchGate provides affiliation with Russia to Ukrainian territories, both for the case of authors (https://www.researchgate.net/profile/Vadim-Kramar-2 or https://www.researchgate.net/profile/Semen-Osipovskiy, where Ukrainian city of Sevastopol is marked as Russian Federation) and papers (Kebkal, Konstantin & Kabanov, Aleksey & Kramar, Oleg & Dimin, Maksim & Abkerimov, Timur & Kramar, Vadim & Kebkal-Akbari, Veronika. (2024). Practical Steps towards Establishing an Underwater Acoustic Network in the Context of the Marine Internet of Things. Applied Sciences. 14. 3527. 10.3390/app14083527, where Ukrainian city of Sevastopol is marked as Russian Federation).

**Study 5: Examples of academic studies which propagate Russian propaganda narratives in support of aggression against Ukraine**

During our investigation, we observed hundreds of cases of spread of Russian propaganda narratives and support of aggression against Ukraine in academic papers published in both Russian and international journals. Some examples are provided in Table 5.

**Table 5. Russian Propaganda in Academic Papers: Selected Cases indexed in Scopus or published by international publishers**

| Journal Title | Example of propaganda |
|---|---|
| Russian Social Science Review, published by Taylor and Francis, UK | A justification of the legality of the Crimea annexation |
| Eurasian Geography and Economics, published by Taylor and Francis, UK | A justification of the Crimea annexation |
| Public Health and Life Environment, published by Federal Center for Hygiene and Epidemiology of the Russian Federation | Donetsk and Lugansk People's Republics, former Donetsk and Lugansk regions), Zaporozhye and Kherson regions are mentioned and discussed as |
| | parts of Russian Federation (found through Scopus) |
| Almanac of Clinical Medicine, Published by EcoVector, Russia | Ukrainian city of Donetsk is mentioned as a part of Russian Federation (found through Scopus) |
| Kutafin Law Review, published by Kutafin Moscow State Law University, Russia | Justifies so called DNR and LNR (found through Scopus) |
| Psychology and Law, published by Moscow State University of Phycology and Education, Russia | So called Donetsk People's Republic is mentioned instead of Donetsk Region of Ukraine (found through Scopus) |
| RUSI Journal, published by Taylor and Francis, UK | A propaganda of the Crimea annexation |

Another example is a yearly ebook entitled "The Territories of the Russian Federation" published by Taylor & Francis which "includes surveys covering the annexed (and disputed) territories of Crimea and Sevastopol" (although they are marked as such). Taylor & Francis reprints in its Journal "Welding International" papers from Russian-language Journal "Сварочное Производство" (Svarochnoe Proizvodstvo), which is promoted on its website to "the defense industries in terms of dynamic rearmament." Taylor & Francis responded to our request for comment: "We cannot comment on T&F's contractual relationships publicly."

Also, there are cases with direct support of Russian aggression against Ukraine and annexation of Ukrainian territories present in the SSRN repository. For example: Burke, John J. A. and Panina-Burke, Svetlana, The Reunification of Crimea and the City of Sevastopol with the Russian Federation: Logic Dictating Borders (June 2, 2017). Available at SSRN: https://ssrn.com/abstract=2979268 or http://dx.doi.org/10.2139/ssrn.2979268 where the following is mentioned (a direct quote): "Crimea and the City of Sevastopol justifiably separated from Ukraine and reunified with the Russian Federation in 2014." Overall hundreds of cases are observed.

## Conclusions

This note highlights how Russia uses the international academic sphere (scientometric databases, international publishers, and international organizations) as a vehicle for legitimization of its appropriation of Ukrainian territories. It is important that the international scientific community put an end to this misuse of its resources. In particular, the academic journals, publishers and scientometric databases should monitor the material published on their websites to prevent the publication, indexing and display of the Russian propaganda related to its appropriation of the Ukrainian territories and to sanction the organizations and authors who produce propaganda. By ensuring responsibility and accountability, the academic community can protect the integrity of scholarly work and uphold the principles of truth and objectivity in research.

## Appendix A

The Ukrainian city of Simferopol is mentioned as part of the Russian Federation by Russian journal “Reviews on Clinical Pharmacology and Drug Therapy” on the site of Scopus (Elsevier)

*Reviews on Clinical Pharmacology and Drug Therapy* • *Открытый доступ* • Том 21, Выпуск 4, Страницы 357 - 366 • 2023

### Tissue oxidative metabolism and microhemodynamics of the skin in rats exposed to stress factors of different durations and combinations

Ravaeva, Marina Yu.[a] ; Cheretaev, Igor V.[a] ; Chuyan, Elena N.[a] ; Galenko-Yaroshevskii, Pavel A.[b] ; Dzheldubaeva, Elviza R.[a] ; Mironyuk, Irina S.[a]

Сохранить всех в список авторов

[a] V.I. Vernadsky Crimean Federal University, Simferopol, Russian Federation
[b] Kuban State Medical University, Krasnodar, Russian Federation

1 48th percentile
Цитата в Scopus

Просмотреть все параметры >

Просмотр PDF ↗ Опции полного текста ∨ Экспорт ∨

https://www.scopus.com/record/display.uri?eid=2-s2.0-85184896742&origin=resultslist&sort=plf-f&src=s&sid=4410ad20f3619bc208c3fbe300355ebd&sot=b&sdt=b&s=DOI%2810.17816%2FRCF609553%29&sl=23&sessionSearchId=4410ad20f3619bc208c3fbe300355ebd&relpos=0

# Appendix B

The Ukrainian city of Sevastopol is mentioned as part of the Russian Federation by Russian journal “Bulletin of the Russian Academy of Sciences. Physics” published by Springer Nature

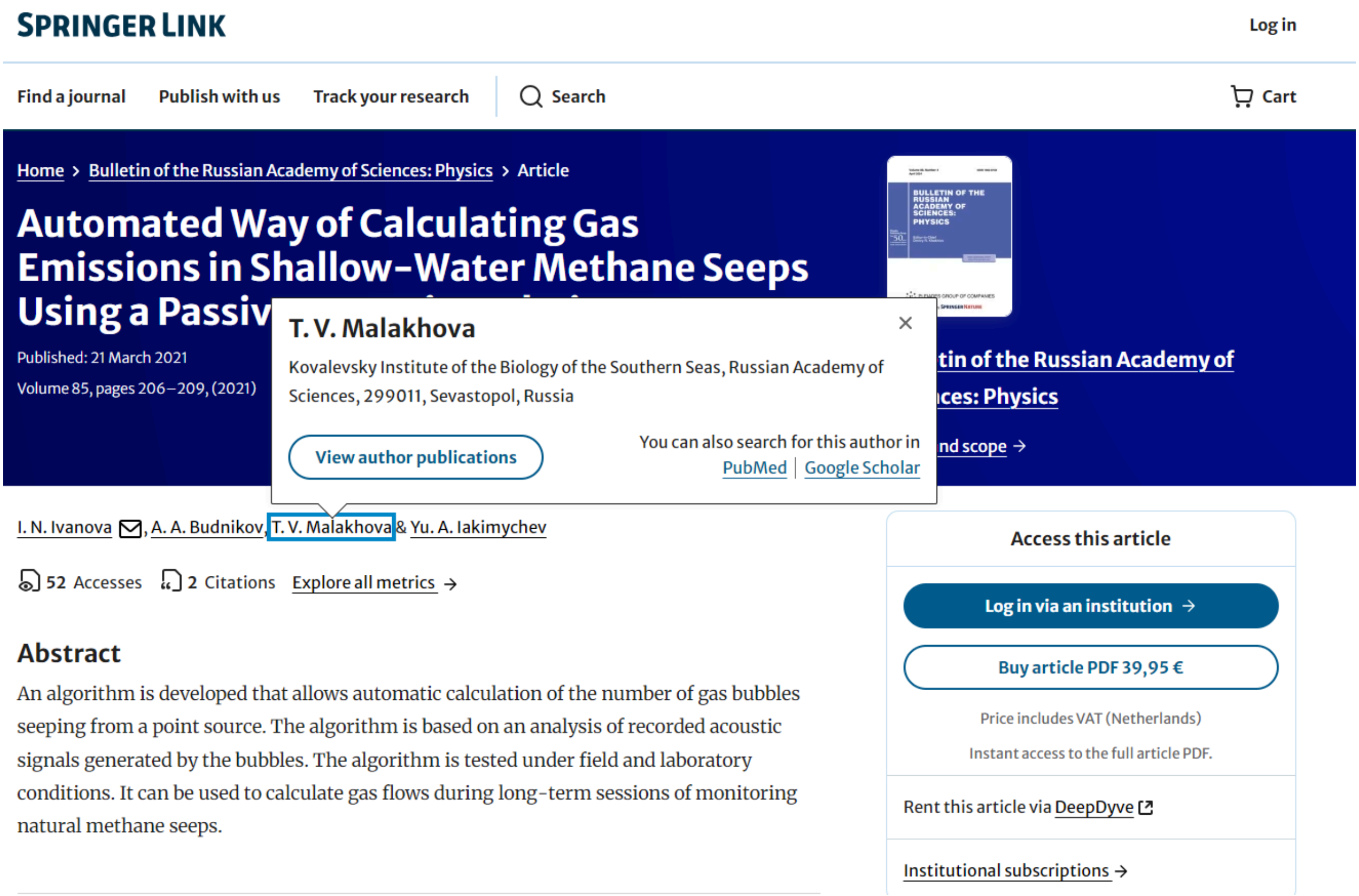

SPRINGER LINK

Log in

Find a journal Publish with us Track your research Search Cart

Home > Bulletin of the Russian Academy of Sciences: Physics > Article

Automated Way of Calculating Gas Emissions in Shallow-Water Methane Seeps Using a Passiv

Published: 21 March 2021

Volume 85, pages 206–209, (2021)

T. V. Malakhova

Kovalevsky Institute of the Biology of the Southern Seas, Russian Academy of Sciences, 299011, Sevastopol, Russia

View author publications

You can also search for this author in PubMed | Google Scholar

tin of the Russian Academy of ces: Physics

nd scope →

I. N. Ivanova, A. A. Budnikov, T. V. Malakhova & Yu. A. Iakimychev

52 Accesses 2 Citations Explore all metrics →

## Abstract

An algorithm is developed that allows automatic calculation of the number of gas bubbles seeping from a point source. The algorithm is based on an analysis of recorded acoustic signals generated by the bubbles. The algorithm is tested under field and laboratory conditions. It can be used to calculate gas flows during long-term sessions of monitoring natural methane seeps.

Access this article

Log in via an institution →

Buy article PDF 39,95 €

Price includes VAT (Netherlands)

Instant access to the full article PDF.

Rent this article via DeepDyve

Institutional subscriptions →

https://link.springer.com/article/10.3103/S1062873821010135

## Appendix C

## The list of 50 Springer Journals spreading the Russian propaganda

| # | Title | Link with example |
|---|---|---|
| 1 | Chemical and Petroleum Engineering | https://link.springer.com/article/10.1007/s10556-018-0385-3 |
| 2 | Applied Magnetic Resonance | https://link.springer.com/article/10.1007/s00723-015-0723-y |
| 3 | Astronomy Reports | https://link.springer.com/article/10.1134/S1063772914120117 |
| 4 | Astrophysical Bulletin | https://link.springer.com/article/10.1134/S1990341319040114 |
| 5 | Atmospheric and Oceanic Optics | https://link.springer.com/article/10.1134/S1024856023060052 |
| 6 | Automation and Remote Control | https://link.springer.com/article/10.1134/S0005117919120099 |
| 7 | Biochemistry (Moscow) | https://link.springer.com/article/10.1134/S000629792008012X |
| 8 | Bulletin of Experimental Biology and Medicine | https://link.springer.com/article/10.1007/s10517-022-05381-x |
| 9 | Bulletin of the Russian Academy of Sciences: Physics | https://link.springer.com/article/10.3103/S1062873818030346 |
| 10 | Colloid Journal | https://link.springer.com/article/10.1134/S1061933X22040111 |
| 11 | Contemporary Problems of Ecology | https://link.springer.com/article/10.1134/S1995425524700161 |
| 12 | Differential Equations | https://link.springer.com/article/10.1134/S0012266123040067 |
| 13 | Doklady Biochemistry and Biophysics | https://link.springer.com/article/10.1134/S1607672922010045 |
| 14 | Doklady Biological Sciences | https://link.springer.com/article/10.1134/S0012496621060053 |
| 15 | Doklady Mathematics | https://link.springer.com/article/10.1134/S1064562424701965 |
| 16 | Entomological Review | https://link.springer.com/article/10.1134/S0013873819080189 |
| 17 | Eurasian Soil Science | https://link.springer.com/article/10.1134/S1064229322700053 |
| 18 | Geochemistry International | https://link.springer.com/article/10.1134/S0016702922020082 |
| 19 | Geotectonics | https://link.springer.com/article/10.1134/S0016852118040027 |
| 20 | Glass Physics and Chemistry | https://link.springer.com/article/10.1134/S1087659622600223 |
| 21 | High Energy Chemistry | https://link.springer.com/article/10.1134/S0018143923010137 |
| 22 | Human Physiology | https://link.springer.com/article/10.1134/S0362119717030057 |
| 23 | Inland Water Biology | https://link.springer.com/article/10.1134/S1995082923050085 |
| 24 | Inorganic Materials: Applied Research | https://link.springer.com/article/10.1134/S2075113316020076 |
| 25 | Journal of Analytical Chemistry | https://link.springer.com/article/10.1134/S106193482305012X |
| 26 | Journal of Applied and Industrial Mathematics | https://link.springer.com/article/10.1134/S1990478919030165 |
| 27 | Journal of Applied Mechanics and Technical Physics | https://link.springer.com/article/10.1134/S0021894418040041 |
| 28 | Journal of Evolutionary Biochemistry and Physiology | https://link.springer.com/article/10.1134/S0022093019060103 |
| 29 | Journal of Ichthyology | https://link.springer.com/article/10.1134/S003294522401003X |
| 30 | Journal of Mathematical Sciences | https://link.springer.com/article/10.1007/s10958-024-06902-x |
| 31 | Journal of Mining Science | https://link.springer.com/article/10.1134/S1062739116010137 |
| 32 | Journal of Russian Laser Research | https://link.springer.com/article/10.1007/s10946-019-09797-1 |
| 33 | Journal of Structural Chemistry | https://link.springer.com/article/10.1134/S0022476619030193 |
| 34 | Journal of Surface Investigation X-Ray, Synchrotron and Neutron Techniques | https://link.springer.com/article/10.1134/S1027451024020204 |
| 35 | Journal of Volcanology and Seismology | https://link.springer.com/article/10.1134/S0742046320030033 |
| 36 | Kinetics and Catalysis | https://link.springer.com/article/10.1134/S002315842401004X |
| 37 | Lobachevskii Journal of Mathematics | https://link.springer.com/article/10.1134/S199508022105022X |
| 38 | Mathematical Notes | https://link.springer.com/article/10.1134/S0001434623110378 |
| 39 | Measurement Techniques | https://link.springer.com/article/10.1007/s11018-016-0893-5 |
| 40 | Mechanics of Solids | https://link.springer.com/article/10.3103/S0025654423070105 |
| 41 | Metallurgist | https://link.springer.com/article/10.1007/s11015-023-01483-7 |
| 42 | Molecular Biology | https://link.springer.com/article/10.1134/S0026893324010114 |
| 43 | Moscow University Biological Sciences Bulletin | https://link.springer.com/article/10.3103/S0096392522040058 |
| 44 | Moscow University Chemistry Bulletin | https://link.springer.com/article/10.3103/S0027131421030044 |
| 45 | Moscow University Geology Bulletin | https://link.springer.com/article/10.3103/S0145875223030122 |
| 46 | Moscow University Soil Science Bulletin | https://link.springer.com/article/10.3103/S014768742303002X |
| 47 | Neuroscience and Behavioral Physiology | https://link.springer.com/article/10.1007/s11055-023-01353-4 |
| 48 | Paleontological Journal | https://link.springer.com/article/10.1134/S0031030120080079 |
| 49 | Pharmaceutical Chemistry Journal | https://link.springer.com/article/10.1007/s11094-020-02166-2 |
| 50 | Physical Mesomechanics | https://link.springer.com/article/10.1134/S1029959914030047 |